# Colossal magnetoresistance over a wide temperature range in Eu$_{0.99}$La$_{0.01}$TiO$_3$

Km Rubi and R. Mahendiran[1]

2 Science Drive 3, Physics Department, Faculty of Science,

National University of Singapore, Singapore 117551

**Abstract**

We report magnetization ($M$), electrical resistivity ($\rho$) and magnetoresistance ($MR$) in polycrystalline Eu$_{0.99}$La$_{0.01}$TiO$_3$. While paramagnetic to antiferromagnetic transition occurs at $T_N$ = 5.46 K upon cooling, $\rho(T, H = 0\,\text{T})$ shows a non-metal to metal transition accompanied by a broad maximum around 65 K in the paramagnetic state. The broad maximum decreases in magnitude and its position shifts to higher temperature with increasing strength of the magnetic field. The magnitude of negative $MR$ at 2.5 K is as large as 42 % in a field of $\mu_0 H$ = 0.6 T and it increases to 75 % for $\mu_0 H$ = 7 T. Though the magnitude of $MR$ decreases with increasing temperature, it is 20 % even at 50 K which is about 10 $T_N$. $MR$ for $T \gg T_N$ nicely fit to the formula $MR = -a^2 \ln(1 + b^2 H^2)$ over a wide field range whereas $MR$ at the lowest temperature scales with $M$. We suggest that electrons doped into Ti-3d band are strongly exchange coupled to localized 4f$^7$ spins of Eu$^{2+}$ ions via f-d interaction. The negative $MR$ is suggested to arise from the field-induced suppression of 4f spin fluctuations and subsequent reduction of scattering of 3d electrons. This is an unique example in perovskite oxides where the magnetoresiststance of 3d electrons is determined by spin fluctuations associated with a rare earth 4f moment. Our results may motivate investigation of magnetoresistance effect which has been overlooked in rare- earth titanates.

---

[1] Corresponding Author (Email: phyrm@nus.edu.sg)

# 1. Introduction

The perovskite oxide EuTiO$_3$ is unique among rare earth titanates of the formula RTiO$_3$ where R = La, Pr, Gd, etc., because it possesses divalent Eu and tetravalent Ti cations[1] contrary to trivalent state adopted by both R and Ti cations in RTiO$_3$ (R = La, Pr, Gd, etc) family.[2,3] At room temperature EuTiO$_3$ has cubic structure with a lattice constant of $a$ = 3.905 Å but it undergoes structural transition (cubic to tetragonal transition similar to 105 K transition in SrTiO$_3$) between 230 K and 282 K depending on oxygen stoichiometry.[4,5] The combination of magnetically active Eu$^{2+}$(4f$^7$) and ferroelectric active Ti$^{4+}$(d$^0$) cations in EuTiO$_3$ is provocative for magnetoelectric interaction. Indeed, bulk EuTiO$_3$ is a G-type antiferromagnetic insulator with a Neel temperature of $T_N$ = 5.4±0.1 K.[6,7] The magnetic phase transition is driven by antiferromagnetic superexchange between 4f$^7$ spins on nearest neighbor Eu$^{2+}$ ions. However, spontaneous ordering of electrical dipoles (ferroelectricity) in bulk EuTiO$_3$ is thwarted down to 2 K by quantum fluctuations and hence EuTiO$_3$ is considered as a quantum paraelectric antiferromagnet.[8] Nevertheless, EuTiO$_3$ exhibits magneto-electric coupling as evidenced by a steep decrease of the static dielectric constant ($\varepsilon$) at $T_N$ in the absence of external magnetic field and a large positive magnetodielectric effect accompanying spin flop transition in the antiferromagnetic state ($\Delta\varepsilon/\varepsilon$ = +7% for $\mu_0 H$ = 7 T and $T$ = 2 K).[8] Interestingly, ferroelectricity and ferromagnetism can be induced by tensile stress in EuTiO$_3$ epitaxial thin film grown on DyScO$_3$ substrate.[9] Ferromagnetism but not ferroelectricity is induced in bulk EuTiO$_3$ due to electron doping either by heterovalent cation substitution at the A-site (La$^{3+}$ for Eu$^{2+}$)[10] or at B-site (Ti$^{4+}$ by Nb$^{5+}$)[11,12] or by oxygen deficiency.[13] The ferromagnetic Curie temperature ($T_C$) in Eu$_{1-x}$La$_x$TiO$_3$ series attains a maximum value for $x$ = 0.1($T_C$ = 8 K) whose the zero-field resistivity shows a metallic behavior but with a pronounced peak at $T_C$.[10] Application of an external magnetic field diminishes the amplitude of the resistivity peak resulting in -24% magnetoresistance at $T_C$ in a field of 7 Tesla.[10]

In contrast to colossal magnetoresistive manganites or cobaltites where 3d band is the source of both magnetism and electrical conduction,[14] magnetism in EuTiO$_3$ is dominated by localized 4f electrons of Eu$^{2+}$ whereas the electrical transport is governed by charge carrier motion in the Ti-3d band. The most interesting situation will be magnetic interaction of low-density electrons in 3d band with localized 4f spins on the magnetotransport. However, there is no report so far on the influence of magnetic field on electrical resistivity in pure EuTiO$_3$ or Eu$_{1-x}$La$_x$TiO$_3$ for $x \ll 0.05$. Here, we report occurrence of 75 % negative magnetoresistance in Eu$_{0.99}$La$_{0.01}$TiO$_3$ which is antiferromagnet. Interestingly, large magnetoresistance exists up to 10 $T_N$ in the paramagnetic state of La substituted sample, which will be interesting for practical applications. Thus, lightly doped EuTiO$_3$ is a new addition to the family of colossal magnetoresistive oxides.

**2. Experimental Details**

Polycrystalline Eu$_{1-x}$La$_x$TiO$_3$ ($x$ = 0.0, 0.01) samples were synthesized by solid-state reaction method. Powder of Eu$_2$O$_3$, La$_2$O$_3$ and TiO$_2$ were mixed in the stoichiometric ratio. Prior to mixing, La$_2$O$_3$ was preheated at 900° C for 8 hours to remove hydroxides and carbonates. After mixing and grinding, the powder was sintered at 1200°C for 24 hours twice under a reducing atmosphere (95% Ar and 5% H$_2$) which converts Eu$^{3+}$ into Eu$^{2+}$. After sintering, the powder was ground again, pressed in a uniaxial press into a disc shaped pellet and the pellet was sintered at 1300°C for 24 hours in the reducing atmosphere. Structural, magnetic and magnetocaloric properties of these compounds were reported by us in earlier publications.[15,16] The sample is single phase and cubic at room temperature. Temperature and field dependences of four probe dc resistivity and magnetization were measured using a Physical Property Measurement System (PPMS) and PPMS based vibrating sample magnetometer (VSM), respectively.

**3. Results and Discussion**

In the main panel of Fig. 1(a) we compare the temperature dependence of magnetization, $M(T)$, of Eu$_{1-x}$La$_x$TiO$_3$ ($x$ = 0.0 and 0.01) measured while cooling under a magnetic field of $H$ = 100 Oe. Only the low temperature data is shown for clarity. The parent compound ($x$ = 0.0) shows a sharp peak around $T = T_N$ = 5.54 K which signals a transition from paramagnetic to antiferromagnetic state. The $x$ = 0.01 sample exhibits enhanced magnetization over $x$ = 0.0 whereas the Neel temperature is nearly unaffected ($T_N$ = 5.46 K). The inset of Fig. 1(a) compares $M(H)$ at $T$ = 2.5 K. $M(H)$ of $x$ = 0.01 increases nearly linearly with increasing magnetic field up to $\mu_0 H$ = 0.1 T, changes slope then, and shows a weak field dependence above 1 T. For fields $\mu_0 H$ < 1 T, while $M(H)$ curve of $x$ = 0.01 is overall similar to the parent compound, it shows enhanced magnetization value for $\mu_0 H$ < 1 T. At the highest field, $M$ = 6.65 $\mu_B$/f.u. which is slightly lower than 7 $\mu_B$/Eu expected for saturation of Eu$^{2+}$:4f$^7$ moments.

The main panel of Fig. 1(b) shows the temperature dependence of the four-probe dc resistivity, $\rho(T)$, for $x$ = 0.0 and 0.01 in zero external magnetic field. $\rho(T)$ of $x$ = 0.0 increases monotonically with lowering temperature and exceeds 5 M$\Omega$ cm below 50 K and it is not possible to reliably measure the four probe resistivity below 50 K using the PPMS. On the other hand, $\rho(T)$ of $x$ = 0.01 increases with lowering temperature down to ~70 K and goes through a broad maximum around 65 K before decreasing on the low temperature side. Although values of room temperature resistivity of these two samples are comparable, resistivity of the La-doped sample is apparently reduced by more than 5 orders of magnitude at 2 K compared to the parent compound. The inset shows ln$\rho(T)$ versus 1/$T$ for $x$ = 0.01. In high temperature region ($T$ > 350 K), the resistivity follows Arrhenius behavior $\rho(T) = \rho_0 e^{-E_a/k_B T}$, where $E_a$ is the activation energy for hopping to next nearest neighbor and it is 285 meV for $x$ = 0.01. We tried fitting the resistivity of $x$ = 0.01 sample below 300 K with small polaron, large polaron and variable range hopping models but none of them could fit the data satisfactorily over a wide temperature range.

Fig. 2(a) shows $M(T)$ of $x$ = 0.01 and Fig.2(b) shows $\rho(T)$ measured under various values of external magnetic field ($H$). The peak at the Neel temperature in $M(T)$ disappears as the field increases above $\mu_0 H$ = 0.6 T due to transformation of antiferromagnetic state into field-induced ferromagnetic state. $\rho(T, H = 0)$ shows a broad maximum around $T_{MIT} = 65$ K ($> T_N$) and $T_{MIT}$ shifts towards higher temperature accompanied by reduction in the peak value of resistivity ($\rho_{peak}$) at $T_{MIT}$ with increasing strength of the external magnetic field. Temperature dependences of $T_{MIT}$ and $\rho_{peak}$ on the magnetic field are plotted in the inset of Fig. 2(b). The value of resistivity at the lowest temperature (2.5 K) is extremely sensitive to sub 1T magnetic field, e.g., $\rho(H)/\rho(0) = 0.58$ for $H = 6$ kOe. Magnetoresistance (MR) is calculated using the standard definition $MR = \frac{\rho(H)-\rho(0)}{\rho(0)} \times 100$, where $\rho(H)$ and $\rho(0)$ are the resistivity values in a magnetic field $H$ and $H = 0$ Tesla. The inset in Fig. 2(c) shows the temperature dependence of MR for different values of $H$. At 2.5 K, the negative MR increases from ~ 3% for $H = 1$ kOe to 42% for $H = 6$ kOe and then to 73% at $\mu_0 H = 2$ T. However, MR shows only a marginal increase as $H$ increases from 2 T to 7 T. The magnitude of MR increases rapidly as temperature falls below 20 K for $\mu_0 H < 1$ T. When $\mu_0 H \geq 2$ T, MR decreases smoothly with temperature as it increases from 2 K. What is interesting is the existence of large MR over a broad temperature range much above $T_N$. For example, MR is $-20$% at 50 K for $\mu_0 H = 7$ T, which is comparable to $-24$% MR found in single crystalline $Eu_{0.9}La_{0.1}TiO_3$ around $T_C$ (= 8 K).[10] At 70 K, MR is -6% at 70 K. Occurrence of large MR much above $T_N$ is surprising.

Inverse susceptibility of the colossal magnetoresistive manganite $La_{0.7}Ca_{0.3}MnO_3$, shows deviation from the Curie-Weiss fit for temperatures up to ~1.4 $T_C$, which was attributed to formation of nanometer size ferromagnetic clusters (magnetic polarons) accompanied by lattice distortion.[17] To seek such a possibility in our samples, we plot the temperature dependence of the inverse susceptibility ($\chi^{-1} = H/M$) for different values of $H$ in the main panel of Fig. 2(c). $\chi^{-1}(T)$ curves for different $H$ values show visible deviations from each other for temperatures below ~35 K but the difference is negligible at higher temperature. While $\rho(T)$ at $\mu_0 H = 7$ T shows a maximum around 80 K (see Fig. 2(b)), no anomaly appears in $\chi^{-1}$ (*T*) at this temperature. Hence, the presence of ferromagnetic clusters much above $T_N$ is doubtful in this compound.

Fig. 3 (a) and (b) show the magnetic field dependence of magnetization and magnetoresistance, respectively. *M* increases linearly with *H* for $T \geq 28$ K and deviates from the linearity at lower temperatures. At 2.5 K, *M* shows a rapid increase for fields below 1 T and tendency to saturate at higher magnetic fields. *MR* is negative and its magnitude increases smoothly with increasing field for $T > 28$ K. We can note that MR is greatly enhanced in lower fields as the temperature drops below 28 K. It is vividly illustrated by the field dependence of *MR* at 2.5 K where the *MR* increases as much as 63 % for a field of 1 T but only incremental change for $H \geq 2$ T. Thus, the field dependence of *MR* is closely related to *M(H)*. The *MR* curves are symmetric for negative values of *H* and no hysteresis is observed while cycling the magnetic field in both directions (not shown here).

This is first time negative magnetoresistance of magnitude comparable to that of manganites has been found in a rare earth titanate. Very recently, Ito *et al.*[18] reported a linear positive *MR* behavior in $RTiO_3$ (R = Pr, Ce) whose origin was attributed to Zeeman splitting of Ti-3d conduction band. However, *MR* is negative in our sample. Let us briefly discuss plausible origin of the observed negative magnetoresistance in our sample. $EuTiO_3$ is a band semiconductor with a direct band gap of 3.2 eV that separates the Ti-3d states dominated conduction band from the O-2p states derived valance band. However, a narrow Eu-4f band lies approximately 0.67 - 0.96 eV below the conduction band edge within the forbidden band gap ($E_F$) and the Fermi level is just on the top of the 4f band.[19,20] As $Ti^{4+}$ is a $d^0$ ion, there are no free charge carriers in the Ti-3d band of $EuTiO_3$. While partial substitution of $La^{3+}(4f^0)$ for $Eu^{2+}(4f^7)$ partially dilute antiferromagnetic interaction among $Eu^{2+}$ ions, it also creates donor impurity states that lie very close to, or overlap with the bottom of Ti-3d conduction band. We assume that spin of an electron ($S_{3d} = ½$) in the donor state is strongly exchange coupled to the localized 4f spins ($S_{4f} = 7/2$) of $Eu^{2+}$ ion via f-d exchange interaction. In the absence of an external magnetic field and above room temperature, doped electrons are thermally excited to Ti-3d($t_{2g}$) band. As temperature decreases below room temperature, activation energy for thermal excitation is insufficient to excite electrons to the conduction band and the resistivity increases. However, transition into metallic state occurs below 65 K. This transition could be due to increase in mobility of electrons that were previously excited to Ti-3d band similar to the mobility enhanced insulator-metal transition reported in $BaTiO_3$ doped with Nb [21] or due to shift of Fermi level into the Ti-3d conduction band as the carriers are introduced by La-substitution. Since donor electrons are exchange coupled to Eu:4f moments, they experiences additional magnetic scattering (spin-disorder scattering) due to 4f spin fluctuations in zero magnetic field. As the external magnetic field is applied above $T_N$ and increased in strength, spin fluctuations diminish progressively which reduces spin-disorder scattering of 3d electrons and hence the resistivity decreases smoothly. Interestingly, spin fluctuations seems to persist much above $T_N$ and impacts the resistivity at temperature as high as 70 K (~ 12 $T_N$) which is unusual. Within the antiferromagnetic state, application of external fields destabilizes the antiferromagnetic spin order. The critical field for spin-flop transition in the parent $EuTiO_3$ itself is very small (~ 6 kOe at 2 K) and the rapid increase of magnetization seen for magnetic fields below 1 T in the present compound is due to decrease of relative angle between flopped spins. In this field range, resistivity decreases rapidly. As the field increases above 1 T, neighboring Eu-4f moments are nearly parallel to each other and the donor electron experiences less scattering leading to only an incremental increase in the magnetoresistance value.

Magnetoresistance due to scattering of conduction electrons by localized 3d impurity magnetic moment in dilute magnetic alloy such as $Cu_{1-x}Mn_x$ was predicted by T. Kausya to scale with the square of the reduced magnetization in low magnetic fields i.e., $\Delta\rho/\rho = C(M/M_s)^2$, where $M_s$ is the saturation magnetization and $C$ is proportionality constant. [22] For low magnetic fields, $M \propto H$ and hence $\Delta\rho/\rho \propto H^2$. We could fit our data over a full magnetic field range ($\mu_0 H$ = 0 to 7 T) with the above relation for $T$ > 40 K but deviations from $M^2$ behavior occurred at higher fields with lowering temperature below 40 K. Khosla and Fischer[23] extended the Kausya's model to include third order correction in the s-d interaction Hamiltonian and shown that *MR* over a wide field range can be described by the empirical relation.

$$\frac{\Delta\rho}{\rho} = -a^2 \ln(1 + b^2 H^2) \qquad (1)$$

where

$$a^2 = A_1 JD(E_F)[S(S+1) - \langle M^2 \rangle] \qquad (2)$$

$$b^2 = \left[1 + 4S^2\pi^2 \left(\frac{2JD(E_F)}{g}\right)^4\right] \frac{g^2\mu_B^2}{(\alpha k_B T)^2} \qquad (3)$$

Here, *J* is the exchange interaction integral, *S* is the spin of the localized moments, *g* is the effective Lande's g-factor, $D(E_F)$ is the density of states at the Fermi level, $<M^2>$ is the average magnetization squared and $\alpha$ is a numerical constant of the order of unity. The parameter *A₁* is the measure of the contribution of spin scattering to the total magnetoresistance. For small magnetic fields, the above empirical relation leads to $\Delta\rho/\rho \propto$ $<M^2>$. We could fit the MR data over a wide field range with Eq. (1) and the fits are shown by lines in Fig. 3(b). While Eq. (1) fits the field dependence of *MR* over the full field range for $T \geq 28$ K, high field data at lower temperatures could not be fitted. When the sample is in the antiferromagnetic state (see 5 and 2.5 K data), Eq.(1) fits only for $\mu_0 H < 1.5$ T. Inset of Fig. 3(c) shows the temperature dependence of the *a* and *b* parameters. While *a* shows a rapid decrease below 20 K *b* increases rapidly. The main panel of Fig. 3(c) shows the field dependence of -*MR* and *M* on the left and right y-axis, respectively. It is notable that -*MR* vs *H* closely follows *M(H)*.

It is worth to recall electrical transport in EuO at this point. EuO is a ferromagnetic semiconductor ($T_C$ = 69.8 K) and it shows colossal magnetoresistance around $T_C$ in the presence of oxygen deficiency or excess Eu ion.[24] The conduction band of EuO is made up of Eu-5d and 6s states and the valence band is made up of O-2p state whereas the narrow 4f band lies close to the conduction band edge. Oxygen vacancy introduces shallow donor (impurity) states very close to the conduction band edge. A strong *f-d* exchange interaction between localized Eu:4f electrons and itinerant Eu:5d electrons causes exchange splitting of the conduction band at $T_C$. The spin-up 5d band moves down by 0.6 eV upon entering the ferromagnetic state from the paramagnetic state.[25] Application of an external magnetic field close to $T_C$ causes further movement of spin-up 5d conduction band edge downwards, which then starts to overlap with the impurity states. Those electrons localized in the shallow donor states are emptied into the spin-up 5d band and hence the resistivity decreases dramatically under external magnetic fields. First principle calculations predict overlap of Ti-3d and Eu-5d state and also non negligible overlap of Ti-3d and Eu-4f states.[26] It needs to be verified experimentally whether spin splitting of 5d band is significant in EuTiO$_3$.

## 4. Conclusion

In summary, Eu$_{0.99}$La$_{0.01}$TiO$_3$ undergoes a non-metal to metal transition around 65 K upon lowering temperature in the absence of an external magnetic field and much above the antiferromagnetic transition ($T_N$ = 5.43 K). A large negative magnetoresistance is observed over a wide temperature. The negative magnetoresistance for 7 T field increases from 6 % at 70 K to 75 % at 2.5 K. At 2.5 K, magnetoresistance as large as −63 % occurs in a field of 1 T. We attribute the negative magnetoresistance to decrease in spin-disorder scattering experienced by Ti-3d electrons following the suppression of Eu-4f spin fluctuations by the external magnetic field. Reports on magnetoresistance in rare earth titanates are still scarce. In view of our findings, it will be interesting exploring the impact of *f-d* interaction on magnetoresistance in other rare earth titanates.

**Acknowledgements**: R. M. acknowledges the Ministry of Education, Singapore for supporting this work (Grant no. R144-000-349-112).

**Figure Captions**

**Fig.1**. (a) Main panel: Temperature dependence of magnetization (*M*) for EuTiO$_3$ and Eu$_{0.99}$La$_{0.01}$TiO$_3$ under the field of 100 Oe. Inset: Magnetic field dependence of *M* at the temperature *T* = 2.5 K for both compounds. (b) Main Panel: Temperature dependence of dc resistivity ($\rho$) for *H* = 0. Inset: ln($\rho$) vs 1/*T* curves with linear fit for Eu$_{0.99}$La$_{0.01}$TiO$_3$.

**Fig.2**. Temperature dependence of (a) magnetization (*M*) and (b) four probe dc resistivity ($\rho$) under various magnetic fields. Inset in (b) shows the magnetic field dependence of metal-insulator transition temperature (left y-axis) and maximum value of resistivity (right y-axis). (c) Temperature dependence of the inverse susceptibility ($\chi^{-1}$) in main panel and magnetoresistance (*MR*) in inset.

**Fig.3**. (a) Magnetic field dependence of magnetization at various temperatures and (b) Magnetic field dependence of *MR* at various temperatures fitted with the eq. $MR = -a^2\ln(1 + b^2H^2)$. (c) Magnetic field dependence of -*MR* (left y-axis) and *M* (right y-axis) at 2.5 K. Inset: Temperature dependence of fitting parameters *a* and *b* for the fit used in Fig. 3(b).

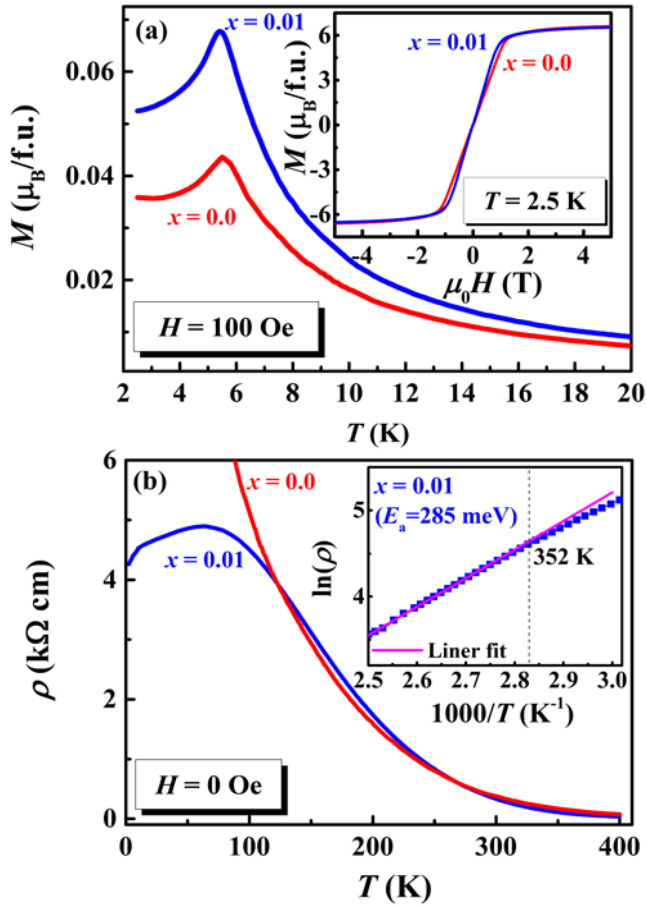

Fig. 1. Km Rubi *et. al.*

**Fig.1**. (a) Main panel: Temperature dependence of magnetization (*M*) for EuTiO$_3$ and Eu$_{0.99}$La$_{0.01}$TiO$_3$ under the field of 100 Oe. Inset: Magnetic field dependence of *M* at the temperature *T* = 2.5 K for both compounds. (b) Main Panel: Temperature dependence of dc resistivity ($\rho$) for *H* = 0. Inset: ln($\rho$) vs 1/*T* curves with linear fit for Eu$_{0.99}$La$_{0.01}$TiO$_3$.

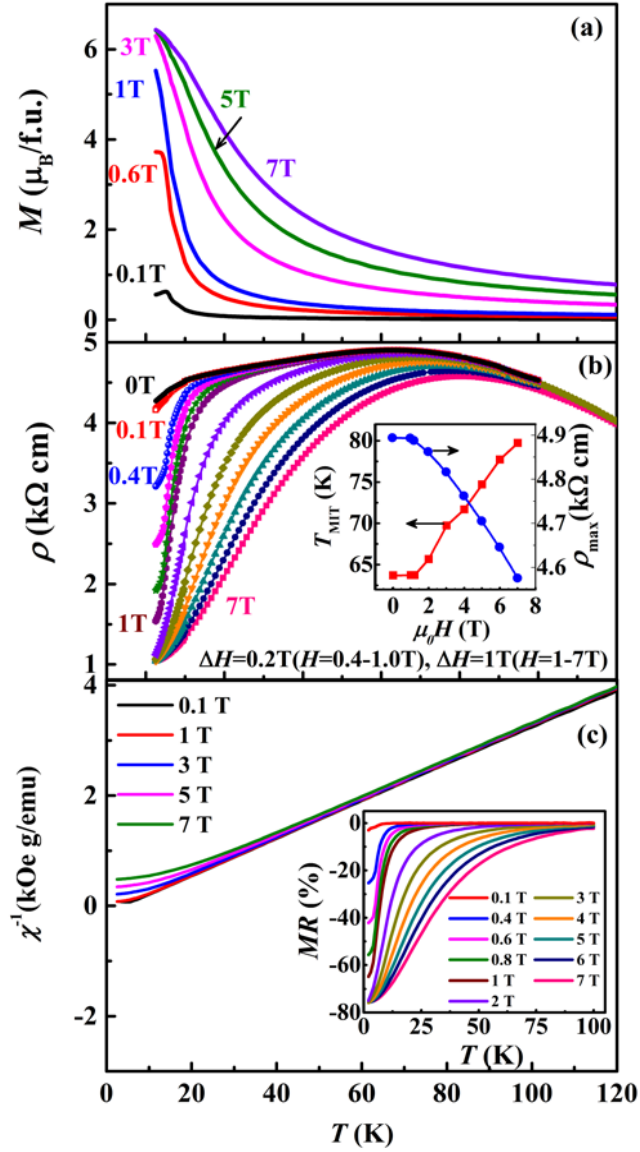

Fig. 2 Km Rubi *et. al.*

**Fig.2**. Temperature dependence of (a) magnetization (*M*) and (b) four probe dc resistivity ($\rho$) under various magnetic fields. Inset in (b) shows the magnetic field dependence of metal-insulator transition temperature (left y-axis) and maximum value of resistivity (right y-axis). (c) Temperature dependence of the inverse susceptibility ($\chi^{-1}$) in main panel and magnetoresistance (*MR*) in inset.

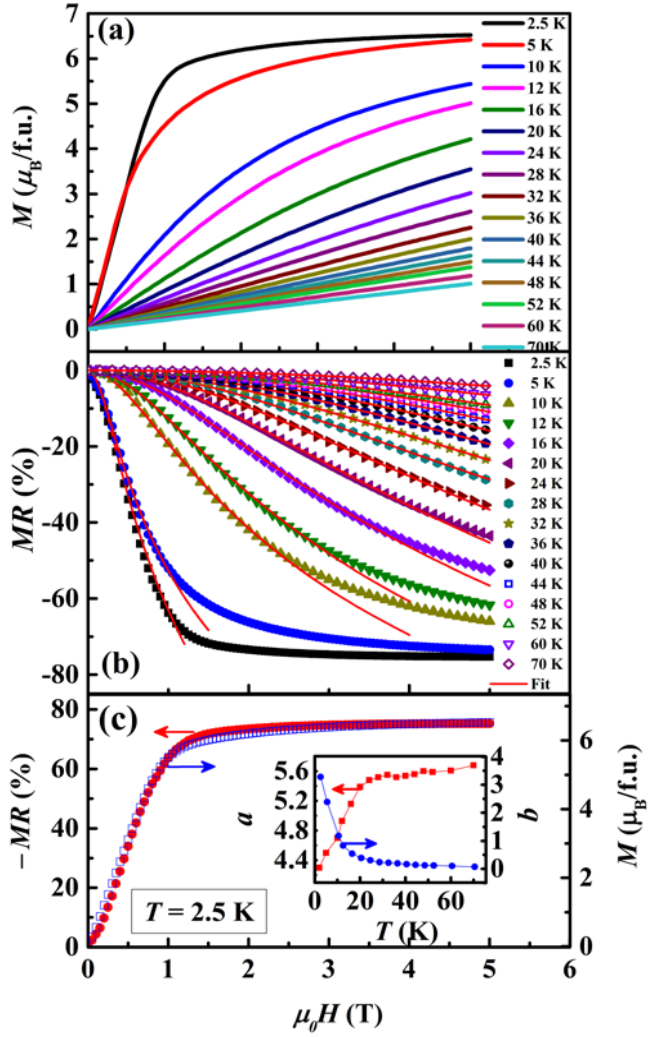

Fig. 3 Km Rubi *et. al.*

**Fig.3**. (a) Magnetic field dependence of magnetization at various temperatures and (b) Magnetic field dependence of *MR* at various temperatures fitted with the eq. $MR = -a^2 \ln(1 + b^2 H^2)$. (c) Magnetic field dependence of -*MR* (left y-axis) and *M* (right y-axis) at 2.5 K. Inset: Temperature dependence of fitting parameters *a* and *b* for the fit used in Fig. 3(b).